\documentstyle[12pt,epsf]{article}

\begin{document}
\baselineskip 18pt
\begin{flushright}
Fermilab-PUB-97/389\\
\today\\
\end{flushright}

\begin{center}
\vspace{0.7 cm}
\large
{\bf{ 
Neutrino Beams from Muon Storage Rings: Characteristics 
and Physics Potential\\
}}
\vspace{1. cm}
\large{S. Geer} \\
\vspace{0.3 cm}
\large{\em {Fermi National Accelerator Laboratory}} \\
\large{\em { P.O. Box 500, Batavia, Illinois 60510}} \\
\end{center}

\vspace{1cm}
\normalsize

\renewcommand{\baselinestretch}{1}
\large
\normalsize

\vspace{1cm}

\abstract{
High-intensity high-energy neutrino 
beams could be produced by exploiting a 
very intense future muon source, and 
allowing the muons to decay in a 
storage ring containing a long straight section. 
Taking the parameters of muon source designs that 
are currently under study, the characteristics 
of the neutrino beams that could be produced 
are discussed and 
some examples of their physics potential given. 
It is shown that the neutrino and antineutrino beam 
intensities may be sufficient to produce hundreds of 
charged current interactions per year in a detector 
on the far side of the Earth.
}

\vspace{1cm}


\vspace{3cm}

Submitted to Physical Review D.
\newpage

\section{Introduction}

High energy neutrino beams have played an 
important role in the development of particle 
physics. Experiments using neutrino and 
antineutrino beams produced from charged meson 
decays have, for example, demonstrated that 
muon neutrinos are different from electron 
neutrinos, discovered neutral currents, provided 
measurements of the structure of the nucleon via 
deep inelastic scattering, made precision tests 
of the Standard Model of electroweak interactions 
via measurements of charged and neutral current 
interactions, and provided increasingly sensitive 
searches for neutrino oscillations in short- and 
long-baseline experiments. 

Recent results from atmospheric neutrino, 
solar neutrino, and short-baseline accelerator 
neutrino experiments indicate that neutrino 
oscillations may occur at rates which are within 
reach of the next generation of accelerator 
based experiments. It therefore seems certain 
that experiments utilizing 
neutrino and antineutrino beams will continue to 
make important contributions to particle 
physics, initially by clarifying whether the 
existing results are indeed indications that 
neutrinos oscillate, and perhaps eventually by 
making precise measurements of the oscillations. 
In addition, there continues to be interest in 
using neutrino beams to further probe the 
structure of the nucleon.

The long-term future of the experimental program 
at neutrino beam facilities will require the 
continued improvement of the intensity and/or quality 
of the beams. The present generation of high-energy 
neutrino beams are made by allowing charged pions 
and kaons to decay in-flight in a long decay channel. 
This paper describes how a new type of neutrino 
beam could be made by exploiting a very high 
intensity muon source of the type that is currently 
being designed as part of the effort to 
develop the technology for a high-luminosity 
muon collider~\cite{snowmass,status}. The muons would be stored and 
allowed to decay in a ring containing a long straight 
section that points in the desired direction. 
The advantages of producing a neutrino beam using muon 
decays are discussed in Section 2. The method of 
producing a neutrino beam from decaying muons is 
described in Section 3. 
The resulting calculated fluxes, which are sufficient 
to provide significant interaction rates in a detector 
on the far side of the Earth, are described in 
Section 4 for short-, long-, and very-long baseline 
neutrino experiments. 
We note that geophysical applications of a facility that 
can shoot neutrino beams through the Earth have been 
discussed in Ref.~\cite{glashow}. 
Some examples of how the neutrino 
beams from a high intensity muon storage ring 
might be used to search for or measure neutrino oscillations 
are given in Section 5, and 
conclusions are summarized in Section 6.

\section{Meson Decays versus Muon Decays}

Presently operating high-energy 
neutrino beams are made by allowing charged pions 
and kaons to decay in a long decay channel.
Shielding downstream of the decay channel removes 
the undecayed mesons, but transmits the weakly 
interacting neutrinos to form a ``pure" neutrino beam. 
If positively charged pions and kaons have been 
selected for the decay channel, the resulting beam 
downstream of the shielding will contain mostly 
muon neutrinos produced in the two-body decays 
$\pi^+ \rightarrow \mu^+\nu_{\mu}$ and 
$K^+ \rightarrow \mu^+\nu_{\mu}$.
The neutrino beam will also contain a small 
component of electron neutrinos produced in the 
three-body decays $K^+ \rightarrow e^+\pi^0\nu_e$. 
In addition, if the primary proton energy is sufficiently 
high, the beam will contain a small $\nu_\tau$ 
component  
coming predominantly from prompt tauonic decays of $D_s$ mesons. 
Antineutrino beams can be made by using a negatively 
charged meson beam. 

Thus, present neutrino (antineutrino) beams consist 
mostly of muon neutrinos (antineutrinos) with a small 
O(1\%) mixture of electron neutrinos (antineutrinos), 
and if the beam energy is high enough, a small component
of tau neutrinos (antineutrinos). 
In general, this beam composition is not ideal for 
neutrino experiments. In particular:
\begin{description}
\item{(i)} The finite precisions with which the 
$\nu_e$ and $\nu_\mu$ fluxes can be 
determined are important sources of systematic 
uncertainty for many neutrino experiments. 
\item{(ii)} The small $\nu_e$ contamination 
in the otherwise pure $\nu_\mu$ beam is a nuisance for 
experiments searching for $\nu_e$--$\nu_{\mu}$ 
oscillations. 
\item{(iii)} The smallness of the $\nu_e$ flux makes 
$\nu_e$--$\nu_{\tau}$ oscillation searches difficult, 
and 
\item{(iv)} The small $\nu_\tau$ contamination in the beam 
will eventually become a nuisance for
experiments searching for $\nu_\mu$--$\nu_{\tau}$
oscillations.
\end{description}

These difficulties can be overcome if the 
neutrino beam is produced by allowing muons to decay 
in the straight section of a storage ring. 
This would produce a beam with a precisely known 
mixture of neutrino types; namely 50\% muon neutrinos 
and 50\% electron antineutrinos if a $\mu^-$ beam is 
stored, and 50\% muon antineutrinos 
and 50\% electron neutrinos if a $\mu^+$ beam is used. 

In the muon rest-frame the distribution of muon 
antineutrinos (neutrinos) from the decay 
$\mu^{\pm} \rightarrow  e^{\pm}$ + $\nu_e$ ($\overline{\nu}_e)$ + 
$\overline{\nu}_{\mu}$ ($\nu_{\mu}$) is given by 
the expression~\cite{gaisser}:
\begin{equation}
\frac{d^2N_{\nu_{\mu}}}{dxd\Omega} = \frac{2x^2}{4\pi}
\left[ (3-2x) \mp (1-2x)\cos\theta \right] ,
\end{equation}
where $x \equiv 2E_\nu$/$m_\mu$, $\theta$ is the angle between 
the neutrino momentum vector and the muon spin direction, and $m_\mu$ is the 
muon rest mass. The corresponding expression describing the 
distribution of electron neutrinos (antineutrinos) is:
\begin{equation}
\frac{d^2N_{\nu_{e}}}{dxd\Omega} = \frac{12x^2}{4\pi}
\left[ (1-x) \mp (1-x)\cos\theta \right] .
\end{equation}
Thus, the neutrino and antineutrino energy-- and angular--
distributions depend upon the parent muon 
energy, the decay angle, and the direction of the 
muon spin vector. 
For an ensemble of muons we must average over the polarization 
of the initial state muons, and the distributions become:
\begin{equation}
\frac{d^2N_{\nu_{\mu}}}{dxd\Omega} \propto \frac{2x^2}{4\pi}
\left[ (3-2x) \mp (1-2x)P_{\mu}\cos\theta \right] ,
\end{equation}
and 
\begin{equation}
\frac{d^2N_{\nu_{e}}}{dxd\Omega} \propto \frac{12x^2}{4\pi}
\left[ (1-x) \mp (1-x)P_{\mu}\cos\theta \right] \; ,
\end{equation}
where $P_{\mu}$ is the average muon polarization along the 
chosen quantization axis, which in this case is the 
beam direction.

The advantages of producing a neutrino beam 
using muon decays rather than meson decays are, therefore:
\begin{description}
\item{(a)} The absolute neutrino fluxes can be easily 
and precisely calculated, provided the 
stored muon current, momentum, and polarization 
are carefully measured. 
\item{(b)} Only one type of neutrino and one type 
of antineutrino are present in the beam, and these types 
can be chosen by selecting the charge of the stored 
muons. Thus precise $\nu_e, \nu_{\mu}, 
\overline{\nu}_e$, and $\overline{\nu}_{\mu}$ measurements 
can be made.
\end{description}
In addition, the muons can be polarized and the time 
dependence of the precessing muon spin vectors monitored, 
enabling measurements to be made as the differential 
spectra of the neutrinos and antineutrinos in the beam 
vary in a precisely known way. 

\section{Using a Muon Storage Ring}

The muon lifetime is about 100 times longer than 
the corresponding charged pion lifetime. For 
example, 20 GeV/c muons have a decay length 
$\gamma$c$\tau = $126~km. Thus, a linear decay 
channel of the type used to produce conventional 
neutrino beams would in practice be too short 
to use efficiently as a muon decay channel. 
This problem can be overcome 
by using a muon storage ring with a straight 
section pointing towards the desired experimental 
area. For simplicity we will consider a 
storage ring that consists of two parallel 
straight sections connected together by two 
arcs. If the straight sections are long compared 
to the arc lengths, the circulating 
muons spend approaching 50\% of their time 
traveling in the straight section pointing 
towards the experiment. In practice there is 
an advantage in keeping the size of the storage 
ring small. Therefore, in the following we will 
assume that the straight sections are equal in 
length to the arcs, and that 25\% of the injected 
muons decay as they circulate in the ring whilst 
they are in the straight section pointing at the 
experiment.

To understand how a muon storage ring designed 
for a given muon momentum might be 
used as a neutrino source, and calculate the 
parameters of the resulting neutrino beams, we 
must understand some of the basic parameters of 
the muon source and storage ring, namely 
the divergence of the beam in the straight sections, 
the size of the ring, 
and the number of muons available from the source. 

If the ring lattice is properly designed, a 
beam divergence $\theta_b \le O(10^{-4})$ should be 
achievable~\cite{carol} in the straight section. 
Thus, if the circulating muons have 
momentum $p/m_{\mu} \ll 10^4$ (corresponding to $p \ll 1000$~GeV/c) 
the angular divergence 
of the neutrino beam produced from decays in the 
straight sections will be dominated by the decay 
kinematics. 

In the scheme we are considering, the size of the 
storage ring is determined by the length of the 
arcs. If the ring is designed to store muons of 
momentum p (GeV/c), the length of each arc in meters 
is given by:
\begin{equation}
L = \frac{\pi p}{0.3 f B},
\end{equation}
where $B$ is the field of the arc dipole magnets (Tesla) 
and the ``packing fraction" $f$ is the fraction of the 
arc lengths occupied by the dipoles. As an example, choosing 
the reasonable values $f = 0.7$ and $B = 8$~T to store 
20~GeV/c muons we obtain 
an estimate of 37~m for the arc lengths and 150~m for the 
ring circumference (two arcs and two straight sections). 
Thus, for muons in this momentum range, it would appear that 
the required storage ring would be sufficiently compact 
to contemplate building the ring in a plane 
tilted at a large angle with respect to the horizon. 
This would enable the neutrino beams to 
be directed through the Earth for a very long baseline 
neutrino oscillation experiment. 
The estimate we are using for the arc lengths has been confirmed by 
a study~\cite{carol_paper} of storage ring 
lattices for rings designed to store muons 
with momenta from 10~GeV/c up to 250~GeV/c. 

The direction of the muon beam must be carefully monitored within 
the straight section to avoid significant systematic uncertainties 
on the calculated neutrino fluxes at the experiment. 
This can be done by placing 
beam position monitors (for example wire chambers) within the muon beam at 
either ends of the straight section. 
The angular precision that would be achieved ($\sigma_\theta$) 
would depend upon the spatial resolution 
of the beam position monitors ($\sigma_x$), and the distance between the 
monitors (L). 
Using Eq.~5 to relate L to the stored muon momentum we obtain:
\begin{equation}
\sigma_\theta \sim \frac{0.3 f B}{\pi p} \; \sqrt{2} \; \sigma_x \; ,
\end{equation}
where $\sigma_x$ is in meters. 
Hence, for B = 8~T, f = 0.7, and choosing $\sigma_x = 500\;\mu$m we obtain 
for p = 20~GeV/c 
the estimate $\sigma_\theta$ = 19~$\mu$r, which is small
compared to the anticipated divergence of the muon beam within the 
straight section. 

The rate at which muons are stored in the ring, 
and therefore the average neutrino beam intensity, will 
be determined by the performance of the muon source. 
In the following we will assume a muon source of the type 
being developed as part of an ongoing effort 
to determine the feasibility of building a high-luminosity 
muon collider~\cite{snowmass,status}. The muon source consists of a 
proton accelerator, 
charged pion production target and collection system, pion 
decay channel, and muon cooling channel. Details of the proton 
accelerator design are given in Ref.~\cite{summer_study}, and 
descriptions of the other components are given in 
Ref.~\cite{snowmass}. For completeness, a brief overview of the 
muon source and its assumed performance is given in the 
following paragraph. 

In the muon collider front-end design that we are taking as an example,
the muon source receives protons from 
an accelerator complex that accelerates bunches containing 
$5 \times 10^{13}$ particles to energies of 16 GeV. 
The  protons interact in a target to 
produce approximately $3 \times 10^{13}$ charged pions of each sign 
per proton bunch. 
These pions are 
produced with only a very limited component of momentum transverse to the 
incident proton direction. The charged pions can therefore 
be confined within a beam channel using, 
for example, a 20~Tesla co-axial solenoid with an inner radius of 7.5~cm. To 
collect as many pions as possible within a useful energy interval, it is 
proposed to use rf 
cavities to accelerate the lower energy particles and decelerate the
higher energy particles. Muons are produced by allowing the pions to decay. 
At the end of a 20~m long decay channel, 
consisting of a 7~Tesla solenoid with a radius of 25~cm, 
on average  0.2 muons of each charge would be produced for each proton 
incident on the pion production target.
With two proton bunches every 
accelerator cycle, the first used to make and collect positive muons and the 
second to make and collect negative muons, 
there would be about $1 \times 10^{13}$ muons of 
each charge available at the end of the decay channel per accelerator cycle. 
If the proton accelerator is cycling at 15~Hz, in an operational year 
($10^7$~secs) about $1.5 \times 10^{21}$ positive and negative muons would 
have been produced in the decay channel and collected.
The muons exiting the decay channel populate a very diffuse phase space.
The next step is to ``cool" the muon bunch, i.e. 
to turn the diffuse muon cloud into a very bright bunch with small 
dimensions in six-dimensional phase-space, suitable for accelerating and 
injecting into a muon storage ring. 
The proposed method of cooling the muons is to use ionization 
cooling~\cite{cooling}. 
At the end of the ionization cooling channel each muon bunch is 
expected to contain about $5 \times 10^{12}$ muons with a momentum of 
order 100~MeV/c. We will assume that the losses in accelerating the 
muons to modest energies (up to a few $\times$~10~GeV or less) 
are small, 
and therefore that $7.5 \times 10^{20}$ muons of the desired charge 
are injected into the storage ring each operational year. 
In the scheme we are considering, 25\% of the muons 
will decay in the straight section pointing at the experimental 
area, and the resulting neutrino beam will contain $2 \times 10^{20}$ 
neutrinos 
per year and $2 \times 10^{20}$ antineutrinos per year, with energy-- 
and angular--distributions described by Eqs. (1) and (2).

\section{Fluxes and Interaction Rates}

In the following we will consider three scenarios, namely 
(i) a very long baseline experiment in which the neutrino 
beam passes through the Earth, (ii) a long baseline 
experiment in which the neutrino beam dips down a few degrees 
with respect to the horizon, and (iii) a ``near" experiment 
which is 1~km from the neutrino source.

\subsection{Very Long Baseline Experiment}

Consider a geometry in which the 
plane of the storage ring dips at an angle of $51^\circ$ 
to the horizon, and the resulting neutrino beam exits the 
Earth at the ``far site" after traversing 9900~km. This geometry would 
correspond to a storage ring sited at the Fermi National 
Accelerator Laboratory in the United States with the far site 
at the Gran Sasso underground Laboratory in Italy. 

\begin{figure}
\epsfxsize6.in
\centerline{\epsffile{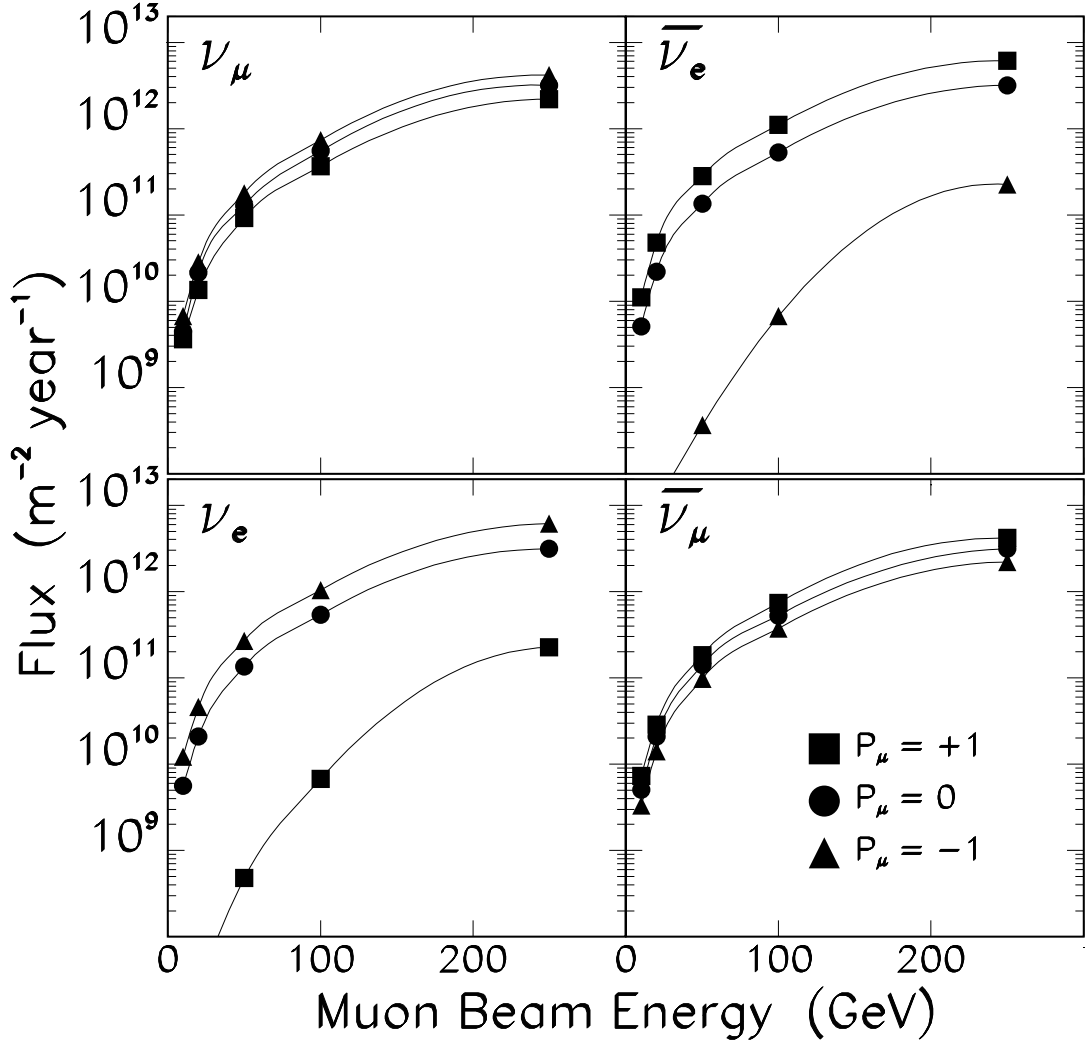}}
\caption{Calculated neutrino and antineutrino fluxes at a far site located 
9900 km from a muon storage ring neutrino source 
(e.g. Fermilab $\rightarrow$ Gran Sasso). 
The parameters of the muon storage ring are described in 
the text. The fluxes are shown as a function of the energy of the stored 
muons for negative muons (top two plots) 
and positive muons (bottom two plots), and for three muon polarizations
as indicated.}
\label{fluxes_fig}
\end{figure}

The calculated neutrino and antineutrino fluxes at the far site 
are shown in Fig.~\ref{fluxes_fig} as a function of the energy 
and average polarization of the muons decaying 
in the straight section of the storage ring. The fluxes have been averaged 
over a 1~km radius ``spot" at the far site. The electron neutrino 
and antineutrino fluxes are very sensitive to the muon spin direction. 
The reason for this can be understood by examining Eq.~2 which 
shows that for $\mu^+$ ($\mu^-$) decays the $\nu_e$ 
($\overline{\nu}_e$) flux $\rightarrow 0$ for all neutrino energies 
as $\cos\theta \rightarrow +1$ ($-1$). 

\begin{figure}
\epsfxsize6.in
\centerline{\epsffile{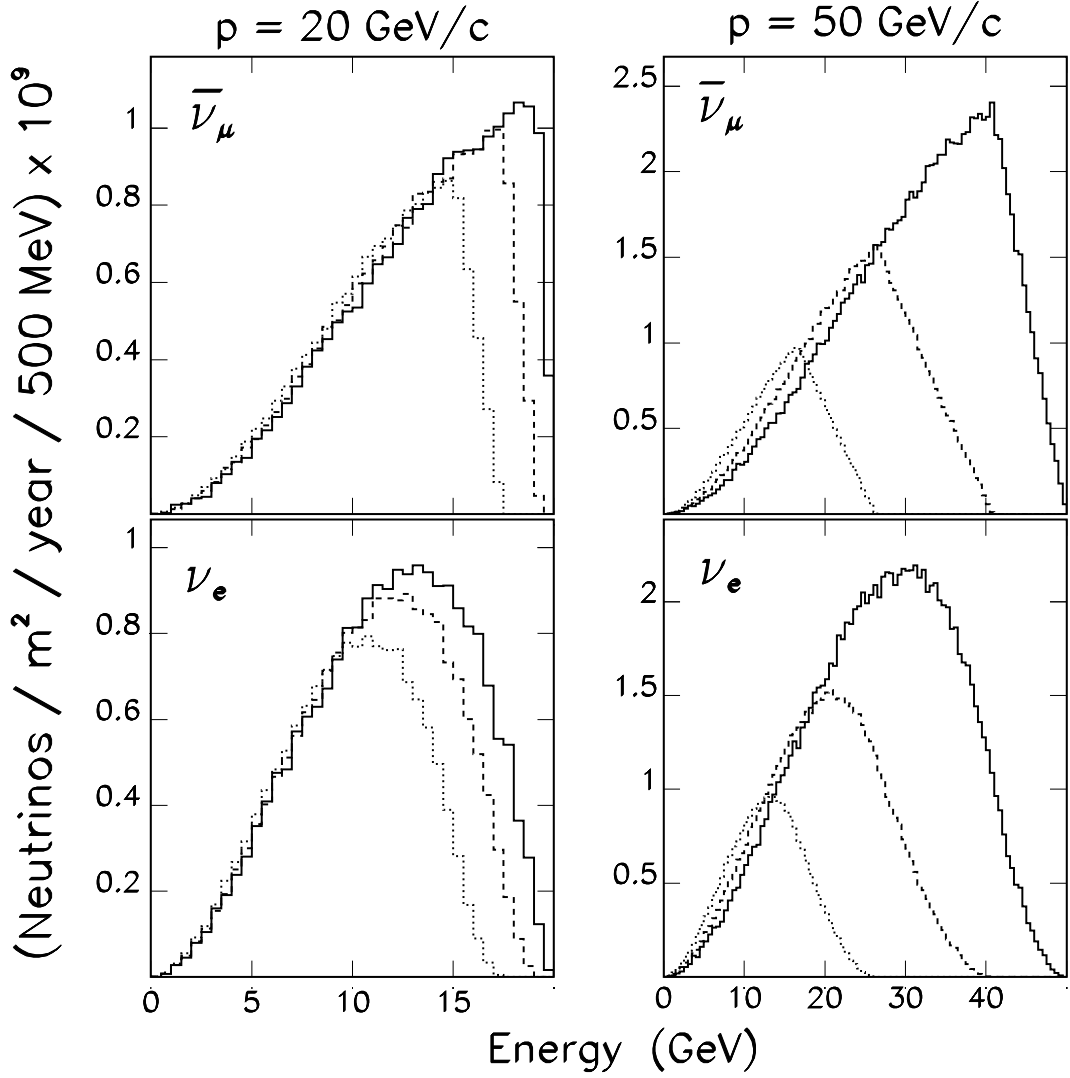}}
\caption{Calculated neutrino and antineutrino differential spectra 
at a far site located 
9900 km from a muon storage ring neutrino source 
(e.g. Fermilab $\rightarrow$ Gran Sasso). 
The parameters of the muon storage ring are described in 
the text. The spectra correspond to unpolarized positive muons circulating 
in the muon storage ring with momenta of 20~GeV/c (left plots) and 50~GeV/c 
(right plots). The solid curves are obtained by averaging the fluxes over 
a central ``spot" with opening angle $\Delta\theta = 1$~mr. The dashed and 
dotted curves are obtained by averaging over annuli centered 
on the beam axis and covering the angular intervals 
$1 < \Delta\theta < 2$~mr and $2 < \Delta\theta < 3$~mr respectively.}
\label{nu_energy_fig}
\end{figure}

As an example we will consider in more detail neutrino beams from 
unpolarized positive muons stored with momenta p = 20~GeV/c (50~GeV/c) 
which, in the absence of neutrino oscillations, produce at the far site 
$2.2 \times 10^{10}$ ($1.4 \times 10^{11}$) 
~$\overline{\nu}_{\mu}$~m$^{-2}$~year$^{-1}$ and 
$2.2 \times 10^{10}$ ($1.4 \times 10^{11}$) 
~$\nu_e$~m$^{-2}$~year$^{-1}$.
These results assume that the neutrino beam is pointing exactly 
in the direction of the far site. The differential distributions 
$dN_{\nu}$/$dE_{\nu}$ are shown in Fig.~\ref{nu_energy_fig} as a function 
of the angle $\Delta\theta$ between the neutrino beam direction and the 
direction of the far site. 
As $\Delta\theta$ increases, both the maximum 
neutrino energy and the neutrino flux decrease. However if, as expected,  
the beam direction can be monitored with a 
precision $\sigma_\theta \ll 1$~mr, the systematic uncertainty on the predicted 
neutrino and antineutrino fluxes and differential distributions 
at the far site should be modest. 

\begin{figure}
\epsfxsize6.in
\centerline{\epsffile{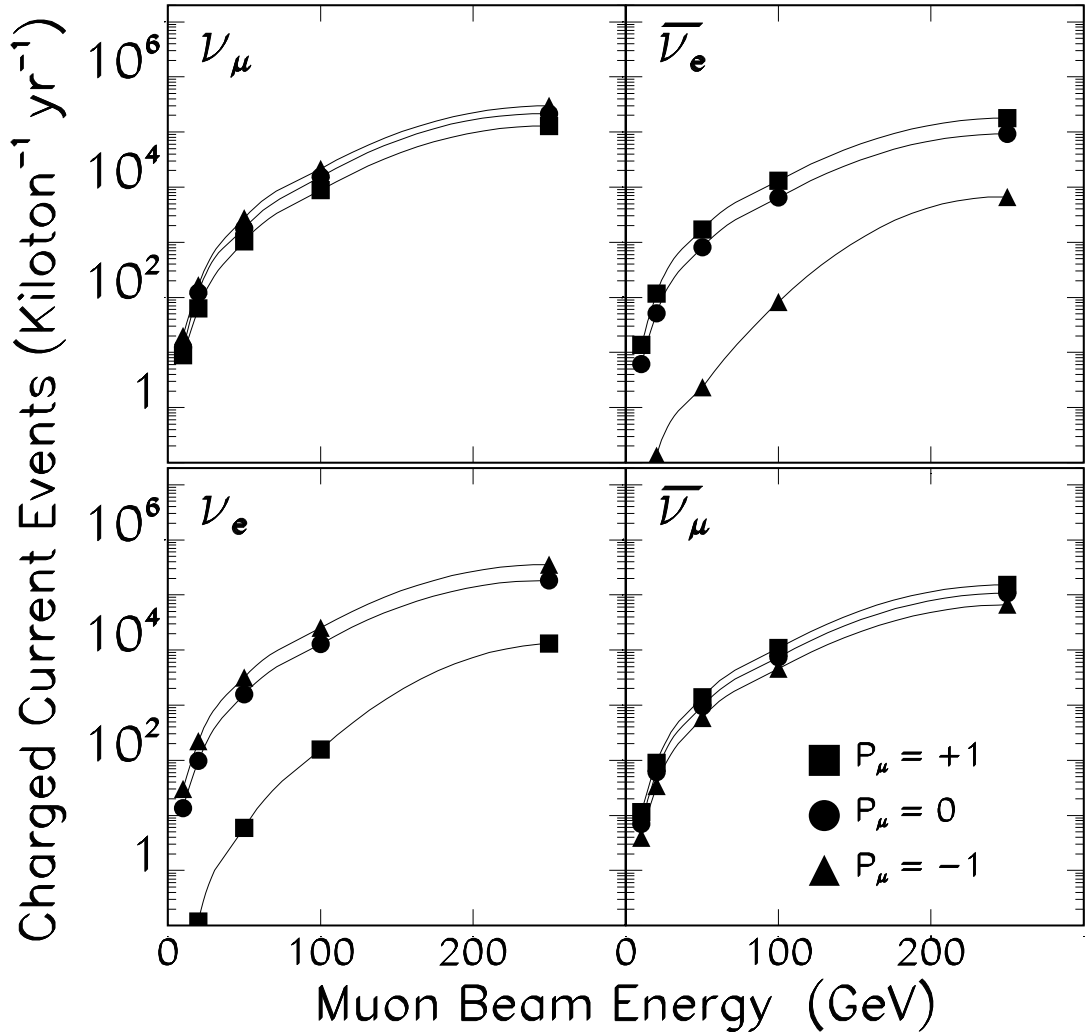}}
\caption{Calculated neutrino and antineutrino charged current 
interaction rates in a detector located 
9900 km from a muon storage ring neutrino source 
(e.g. Fermilab $\rightarrow$ Gran Sasso). 
The parameters of the muon storage ring are described in 
the text. The rates are shown as a function of the energy of the stored 
muons for negative muons (top two plots) 
and positive muons (bottom two plots), and for three muon polarizations
as indicated.}
\label{cc_rates_fig}
\end{figure}

\begin{figure}
\epsfxsize6.in
\centerline{\epsffile{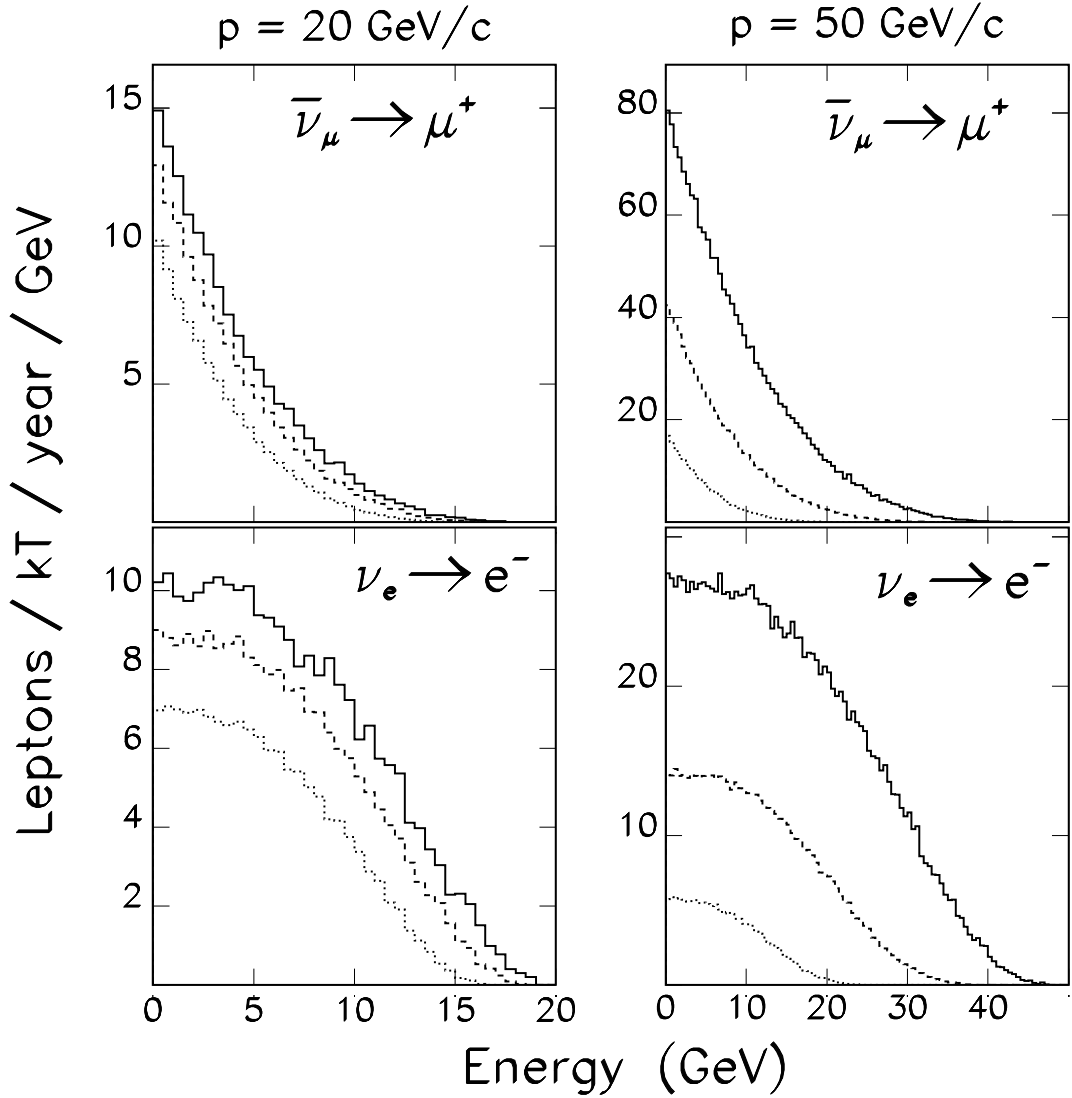}}
\caption{Calculated lepton and antilepton differential spectra 
for particles produced in charged current interactions in a detector located 
9900 km from a muon storage ring neutrino source 
(e.g. Fermilab $\rightarrow$ Gran Sasso). 
The parameters of the muon storage ring are described in 
the text. The spectra correspond to unpolarized positive muons circulating 
in the muon storage ring with momenta of 20~GeV/c (left plots) and 50~GeV/c 
(right plots). The solid curves are obtained by averaging the fluxes over 
a central ``spot" with opening angle $\Delta\theta = 1$~mr. The dashed and 
dotted curves are obtained by averaging over annuli centered 
on the beam axis and covering the angular intervals 
$1 < \Delta\theta < 2$~mr and $2 < \Delta\theta < 3$~mr respectively.}
\label{lepton_energy_fig}
\end{figure}

The charged current neutrino and antineutrino rates in a detector at 
the far site can be calculated using the approximate 
expressions~\cite{boehm} for the cross-sections:
\begin{equation}
\sigma_{\nu N} \;\sim\; 0.67 \times 10^{-38} \; cm^2 \; \times \; E_{\nu} (GeV)
\end{equation}
and
\begin{equation}
\sigma_{\overline{\nu}N} \;\sim\; 0.34 \times 10^{-38} \; cm^2 \; \times \; 
E_{\overline{\nu}} (GeV) \; .
\end{equation}
The predicted charged current interaction rates are shown 
in Fig.~\ref{cc_rates_fig} as 
a function of the energy and average polarization of the decaying muons, 
and the associated charged--lepton 
energy distributions are shown in Fig.~\ref{lepton_energy_fig} 
as a function of the angle $\Delta\theta$ between the beam direction and 
the direction of the far site. 
In the absence of neutrino oscillations, the number of charged current 
interactions in a 10~kT far site detector 
($\Delta\theta$ = 0) when unpolarized 
20~GeV/c (50~GeV/c) positive muons are stored in the ring are 
$610$ ($1.0 \times 10^{4}$) 
~$\overline{\nu}_{\mu}$ interactions per year and 
$1 \times 10^3$ ($1.6 \times 10^{4}$)~$\nu_e$ interactions per year. 
We conclude that, for these particular examples, 
interactions from neutrinos and 
antineutrinos should be readily detectable at the far site.
Note that the predicted charged current interaction 
rates and the shapes of the associated lepton energy spectra are 
both sensitive to $\Delta\theta$. This could be exploited by 
locating one or more satellite detectors at angular distances 
$\Delta\theta$ = O(1~mr) from the main far site detector.

\subsection{Long Baseline Experiment} 

Consider a long baseline geometry in which the plane of the 
muon storage rings tilts at just a few degrees to the horizon. To be 
explicit we will consider a far site that is 732~km from the neutrino 
source. This geometry would 
correspond~\cite{numi} to a storage ring sited at the Fermi National
Accelerator Laboratory and a far site at the Soudan 
underground Laboratory in Minnesota.

The neutrino fluxes and charged current interaction rates at the far 
site can be obtained by scaling the results presented in 
Figs.~\ref{fluxes_fig} and \ref{cc_rates_fig} by a factor of 183. 
Thus, in the absence of neutrino oscillations,  
unpolarized positive muons stored with momenta p = 20~GeV/c (50~GeV/c)
will produce at the far site
$4 \times 10^{12}$~($2.6 \times 10^{13}$) 
$\overline{\nu}_{\mu}$ and $\nu_e$~m$^{-2}$~year$^{-1}$.
In a 10~kT detector, these fluxes would result in 
$1.1 \times 10^5$~($1.8 \times 10^{6}$) 
$\overline{\nu}_{\mu}$ charged current interactions per year and
$1.8 \times 10^5$~($2.9 \times 10^{6}$) 
$\nu_e$ charged current interactions per year.

\begin{figure}
\epsfxsize6.in
\centerline{\epsffile{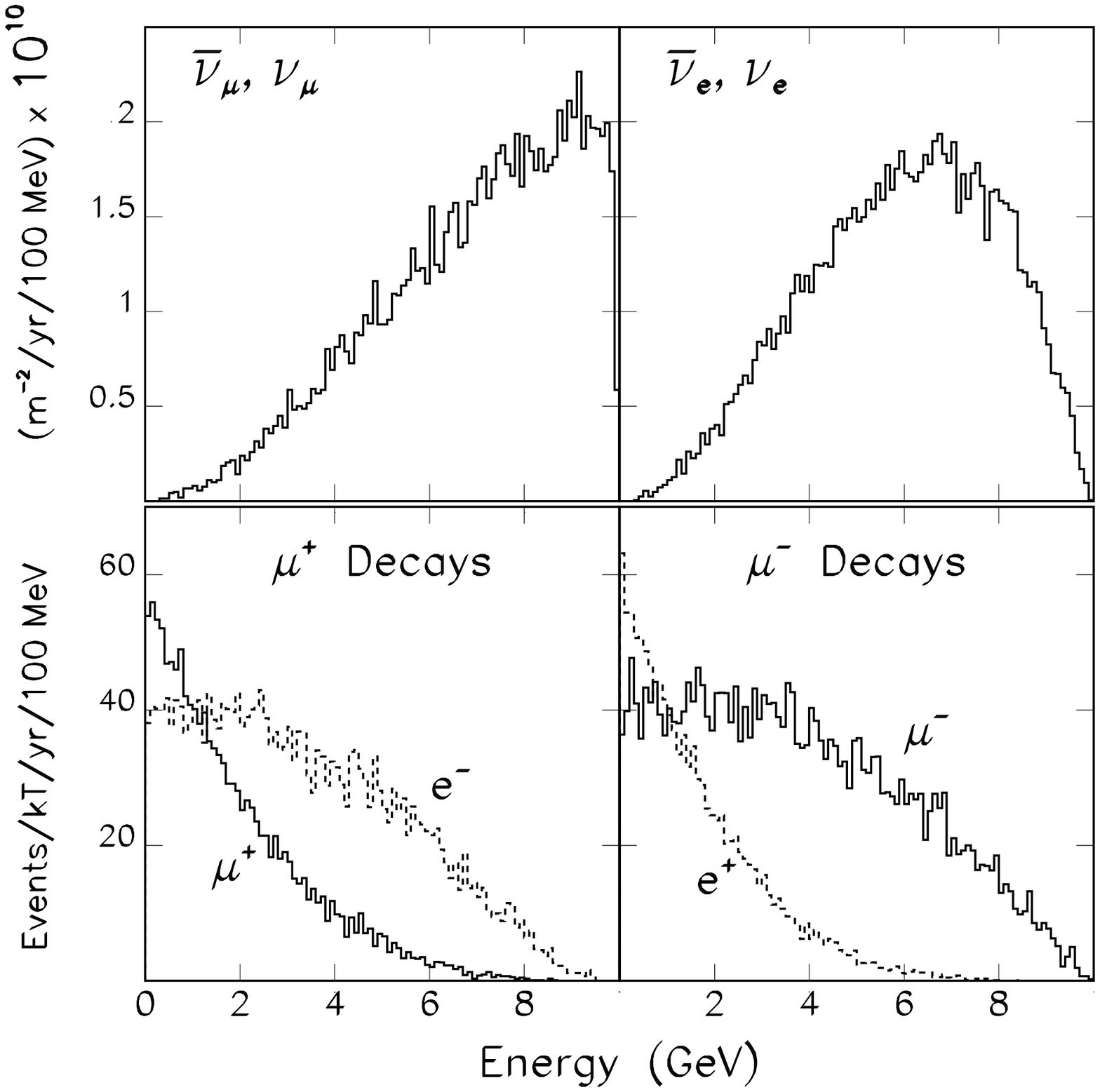}}
\caption{Calculated fluxes and spectra in a detector 732~km downstream 
of a muon storage ring neutrino source in which 10~GeV/c unpolarized 
muons are circulating. The top two plots show the neutrino and 
antineutrino spectra, and the bottom two plots show the charged lepton 
spectra from charged current interactions when positive muons (bottom left)
and negative muons (bottom right) are stored in the ring.}
\label{10gev_fig}
\end{figure}

Given these large interaction rates it is worthwhile considering using a 
lower energy muon storage ring. Predicted fluxes and spectra corresponding 
to using 10~GeV/c stored muons are shown in Fig.~\ref{10gev_fig}. The 
calculated neutrino and antineutrino fluxes are both 
$\sim 1 \times 10^{12}$~m$^{-2}$~year$^{-1}$, and the corresponding 
charged current interaction yields are 
$3.1 \times 10^3$~$\mu^-$~kT$^{-1}$~year$^{-1}$ 
and $1.4 \times 10^3$~$e^+$~kT$^{-1}$~year$^{-1}$ when negative 
muons are stored in 
the ring, and $1.6 \times 10^3$~$\mu^+$~kT$^{-1}$~year$^{-1}$
and $2.9 \times 10^3$~$e^-$~kT$^{-1}$~year$^{-1}$ when positive 
muons are stored in
the ring. The mean energies of the charged leptons and antileptons produced 
in these charged current interactions are respectively $\sim 3.5$~GeV and 
$\sim 2$~GeV. Thus, neutrino and antineutrino charged current 
interactions should be readily detectable at the far site when the decaying 
muons in the storage ring have momenta as low as 10~GeV/c. 

\subsection{Near Experiment}

Next consider a short baseline geometry in which the detector 
is 1~km from the neutrino source. This geometry is of interest to 
short baseline neutrino oscillation experiments if the neutrino 
energies are much lower than those considered so far, and of interest 
to deep inelastic scattering experiments if the neutrino energies 
are much higher than considered so far. 

\begin{figure}
\epsfxsize6.in
\centerline{\epsffile{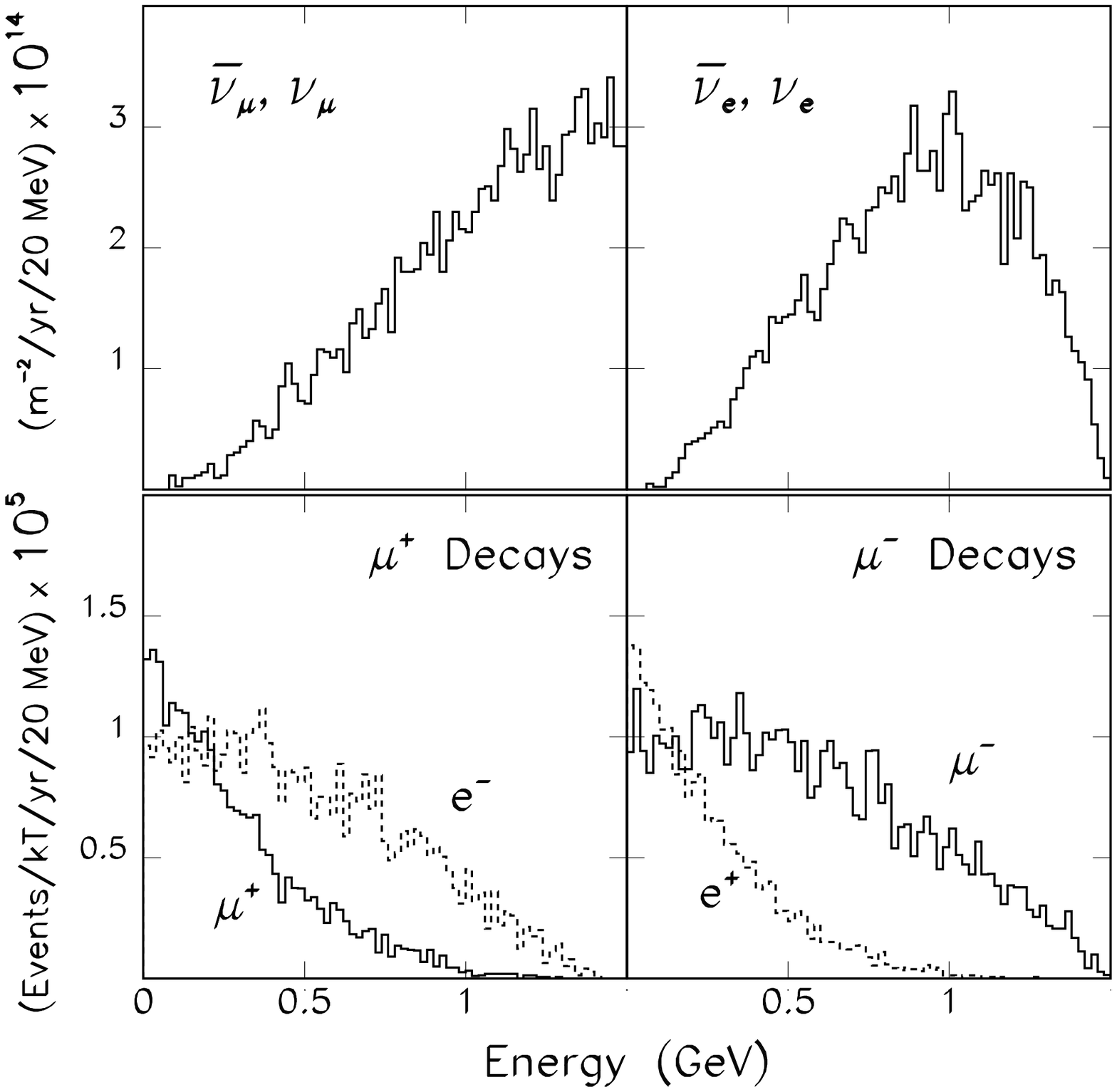}}
\caption{Calculated fluxes and spectra in a detector 1~km downstream 
of a muon storage ring neutrino source in which 1.5~GeV/c unpolarized 
muons are circulating. The top two plots show the neutrino and 
antineutrino spectra, and the bottom two plots show the charged lepton 
spectra from charged current interactions when positive muons (bottom left)
and negative muons (bottom right) are stored in the ring.}
\label{near_fig}
\end{figure}

Consider first a 1.5~GeV/c muon storage ring. 
Averaging the fluxes at the detector over a ``spot" 
with a radius of 5~m, the predicted neutrino and antineutrino 
fluxes resulting from unpolarized muon decays are both 
$1.2 \times 10^{16}$~$m^{-2}$ per year. The corresponding charged 
current interaction rates in a 1~kT detector yield 
$2.9 \times 10^6$~$\mu^+$ per year and $5.1 \times 10^6$~$e^-$ per year 
if positive muons are stored in the ring, and 
$6.0 \times 10^6$~$\mu^-$ per year and $2.6 \times 10^6$~$e^+$ per year
if negative muons are stored in the ring.
The calculated neutrino 
and antineutrino differential spectra are shown in Fig.~\ref{near_fig} 
together with the charged lepton distributions from charged 
current interactions. A 1~kT detector would record millions of 
charged current interactions per year with mean charged--lepton energies of 
$\sim 0.6$~GeV and mean charged--antilepton energies of $\sim 0.3$~GeV. 

Finally we consider a higher energy storage ring suitable for deep inelastic 
scattering experiments. This might be the last recirculating linear accelerator 
ring in a future muon collider accelerator complex. We will assume the muons 
are unpolarized and choose 250~GeV/c 
for the stored muon momentum. After 1~km the neutrino beam 
``spot" has a radius of less than 1~m. Hence nearly all of the O($10^{20}$) 
neutrinos and antineutrinos per year will pass through a reasonably 
compact detector. This, together with the very large neutrino flux, would 
make possible a small detector incorporating detailed tracking and 
particle identification. If, for example, the fiducial mass is 10~kg, 
then the estimated charged current interaction rate is 
$8 \times 10^5$ neutrino interactions 
per year and $5 \times 10^5$ antineutrino interactions per year. 

\section{Examples}

To illustrate the physics potential of the muon storage ring neutrino 
sources discussed in the previous section, consider the sensitivity 
of an experiment searching for $\nu_e$--$\nu_\mu$ or $\nu_e$--$\nu_\tau$ 
oscillations 
performed by searching for charged current interactions producing 
``wrong-sign" muons. For example, if positive muons are stored at the 
neutrino source, unoscillated muon--antineutrino and electron--neutrino 
charged current interactions  
will produce $\mu^+$ and $e^-$ respectively. 
However, if the $\nu_e$ transforms itself into a $\nu_\mu$ during its 
passage to the detector the charged current interaction will produce a 
$\mu^-$. Similarly, if the $\nu_e$ transforms itself into a $\nu_\tau$ 
and the neutrino energy is sufficiently large, 
the charged current interaction will produce a $\tau^-$ which, with a 
branching ratio of 17\%~\cite{pdb}, will decay to produce a $\mu^-$.

Within the framework of two-flavor vacuum oscillations, 
the probability that, whilst traversing a distance L, 
a neutrino of type 1 (mass $m_1$) 
oscillates into a neutrino of type 2 (mass $m_2$) is given 
by~\cite{gaisser}:
\begin{equation}
P(\nu_1 \rightarrow \nu_2) = 
\sin^2(2\theta) \; \sin^2(1.27\Delta m^2 \; L/E) \; ,
\end{equation}
where $\theta$ is the mixing angle, and 
$\Delta m^2 \equiv m_2^2 - m_1^2$ is measured in $eV^2/c^4$, L in km, and 
the neutrino energy E is in GeV. Hence, in the absence of backgrounds or 
systematic uncertainties, a neutrino oscillation  experiment can be 
characterized by the total number of neutrino interactions observed 
(and hence the minimum observable $P(\nu_1 \rightarrow \nu_2)$) and 
the average L/E for the interacting neutrinos. These parameters are 
summarized in Table~1 for the four neutrino oscillation experimental 
configurations discussed in the previous section. 

\begin{figure}
\epsfxsize6.in
\centerline{\epsffile{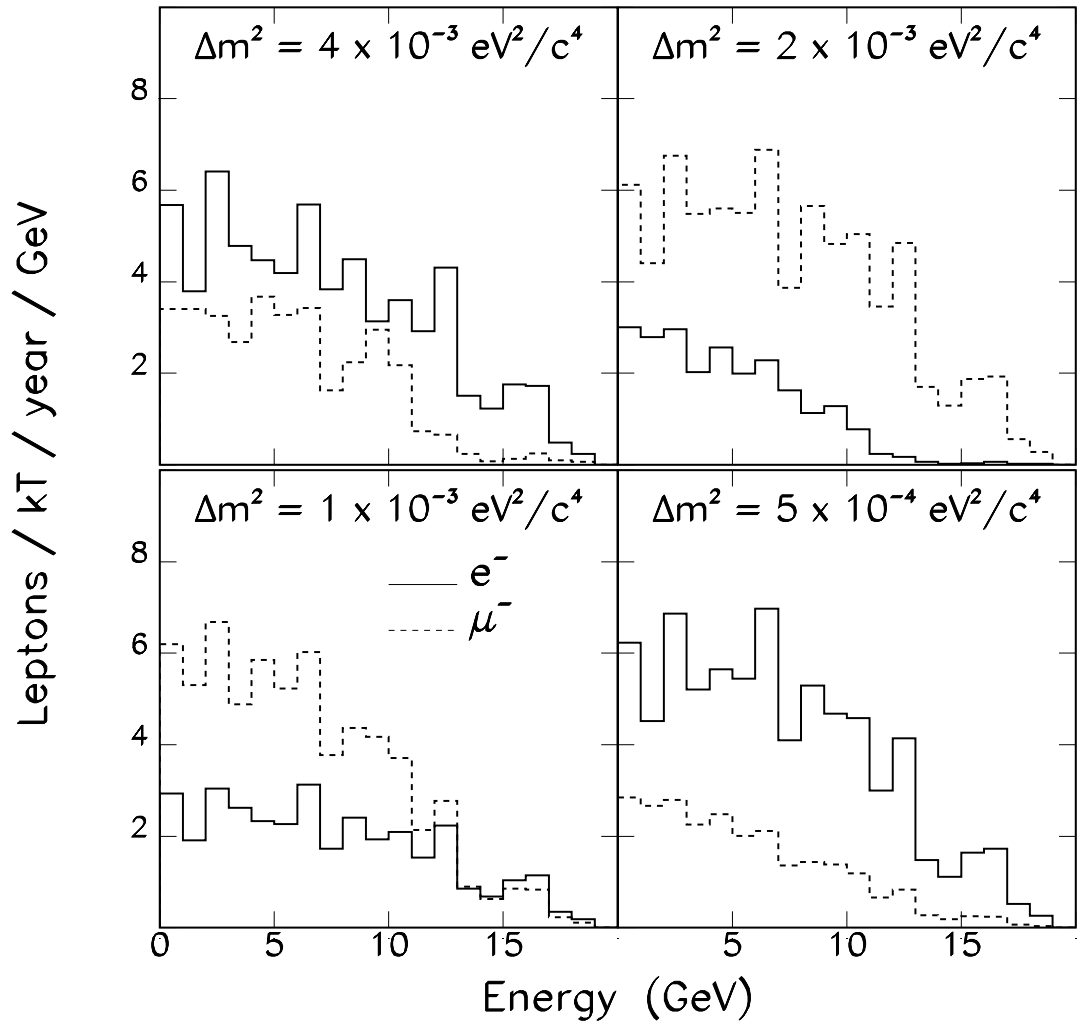}}
\caption{Predicted charged lepton energy distributions in a detector 
9900~km from a ring in which 20~GeV unpolarized positive muons are 
decaying. The predicted differential distributions for 
e$^-$ (solid histograms) and $\mu^-$ 
(broken histograms) are shown assuming 
$\nu_e$--$\nu_{\mu}$ oscillations are occurring with $\sin^22\theta$ = 1 
and $\Delta m^2$ as indicated on the 4 sub-plots.}
\label{oscillated_lepton_energies_fig}
\end{figure}

Consider first the very long baseline (Fermilab $\rightarrow$ Gran Sasso) 
experiment. The mean value of L/E for the interacting $\nu_e$ is 
744~km/GeV, and the minimum observable $P(\nu_e \rightarrow \nu_\mu)$ 
is O($10^{-3}$). 
Examples of the predicted charged current $e^-$ and $\mu^-$ spectra are 
shown in Fig.~\ref{oscillated_lepton_energies_fig} for 
$\sin^2(2\theta) = 1$ and four choices of 
$\Delta m^2$ in the range 
$5 \times 10^{-4} < \Delta m^2 < 4 \times 10^{-3} \; eV^2/c^4$. 
The spectra are clearly sensitive to values of the oscillation parameters 
within this range. 

\begin{figure}
\epsfxsize6.in
\centerline{\epsffile{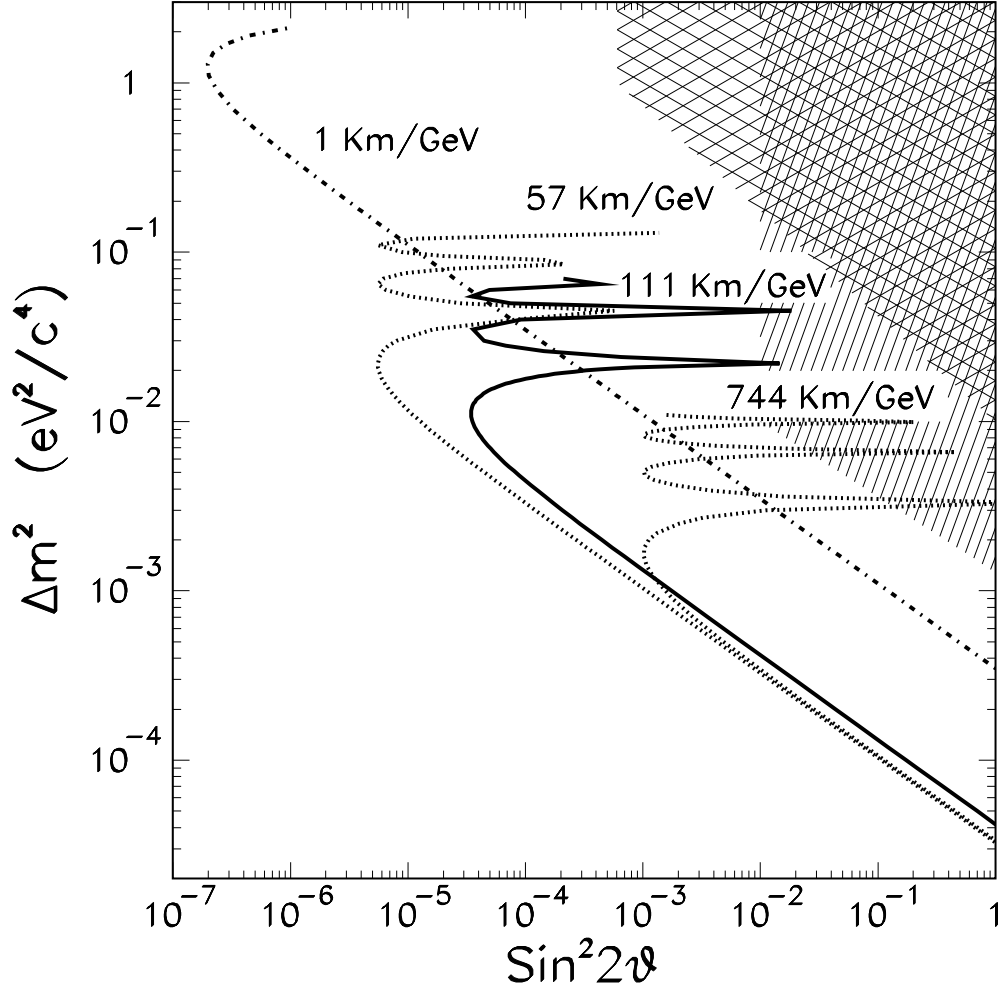}}
\caption{Contours of single--event sensitivity for $\nu_e$--$\nu_\mu$ 
oscillations for 1~year of running with the 4 values of L/E 
specified on the figure, which 
correspond to the 4 detector configurations summarized in Table~1. 
The hatched and cross-hatched areas show the expected regions that will 
be explored by respectively the MINOS experiment~[10] after 2 years of running 
and the MiniBooNe experiment~[11] after 1 year of running.}
\label{plane_fig}
\end{figure}

\begin{figure}
\epsfxsize6.in
\centerline{\epsffile{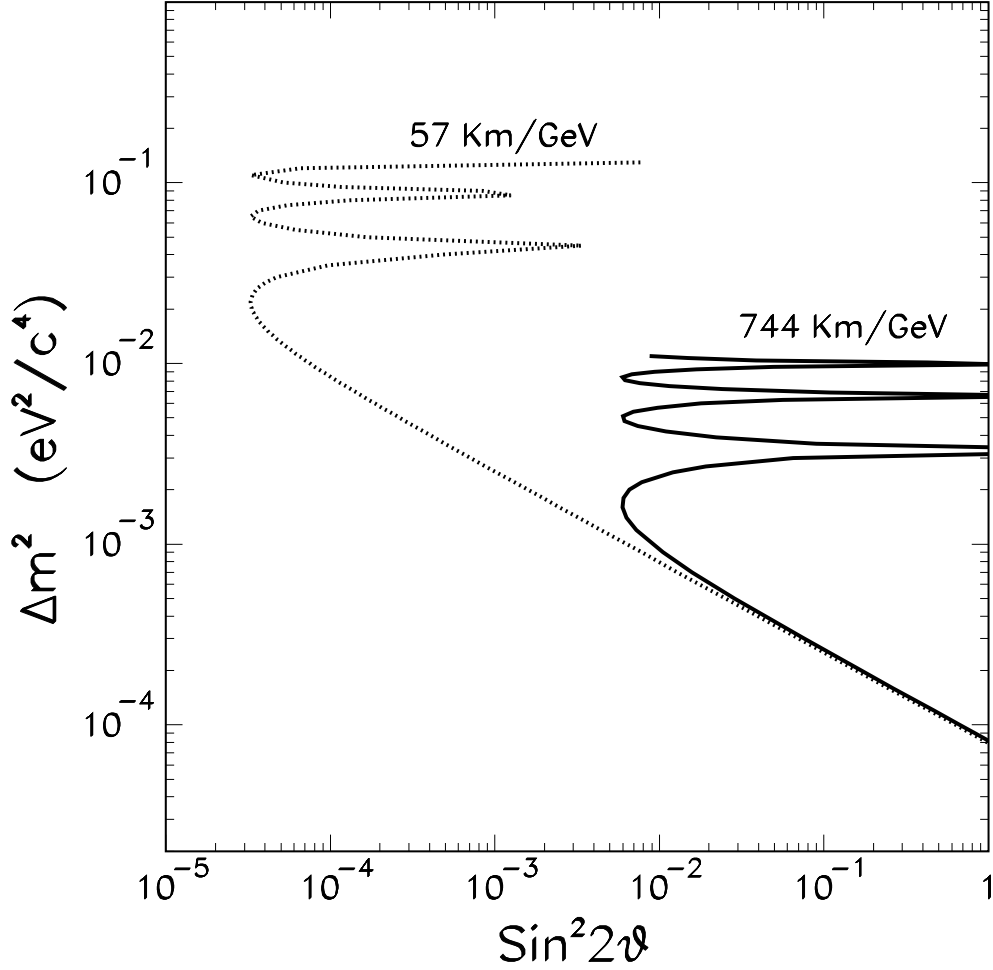}}
\caption{Contours of single--event sensitivity for $\nu_e$--$\nu_\tau$ 
oscillations based on searching for ``wrong-sign" muons with the 
very--long--baseline (solid contour) and long--baseline 
(dotted contour) configurations listed in Table~1 for 
the average L/E values of respectively 744 km/GeV and 57 km/GeV. 
The contours correspond to a 10~kT-year exposure with 20~GeV/c unpolarized 
muons stored in the ring and the straight section pointing at detectors 
9900~km and 732~km from the ring.}
\label{plane_tau_fig}
\end{figure}

Figure~\ref{plane_fig} compares the $\nu_e$--$\nu_\mu$ oscillation 
single--event sensitivity contour in the 
($\Delta m^2$, $\sin^2(2\theta)$)--plane for the very long baseline 
configuration with the corresponding contours for the three other 
configurations discussed in the previous sections. The very long baseline 
single--event contour extends down to 
$\Delta m^2 \sim 3 \times 10^{-5} \; eV^2/c^4$ for $sin^2(2\theta) = 1$, 
with the ``knee" in the contour at $\sin^2(2\theta) \sim 10^{-3}$. 
The predicted single--event contours for the 
two long baseline (Fermilab $\rightarrow$ Soudan) configurations 
have similar $\Delta m^2$ reaches as $\sin^2(2\theta) \rightarrow 1$, 
and have ``knees" at lower values of $\sin^2(2\theta)$ reflecting 
the larger charged current event statistics. 
Note that the $\Delta m^2$ reaches of the long-- and very--long--baseline 
configurations 
are more than an order of magnitude better than 
the expected reaches of the next generation of 
currently proposed neutrino experiments~\cite{numi}--\cite{orlando}. 
The single--event contour for the short baseline configuration exhibits 
a $\Delta m^2$ reach that is only a little better than 
the expected sensitivities of 
the next generation of long-baseline neutrino experiments. 
However, the very large 
charged current event rates expected for the short baseline configuration 
would enable sensitivities approaching 
$\sin^2(2\theta) \sim 10^{-7}$ for large $\Delta m^2$ 
($\Delta m^2 \sim 1 \; eV^2/c^4$). This $\sin^2(2\theta)$ reach 
is almost a factor of 1000 better than the expected reaches of 
presently proposed future experiments, but 
would only be attained in the absence of backgrounds from, 
for example, secondary production of neutrinos from interactions 
in the vacinity of the experiment, or a component of neutrinos 
produced from the decays of ``wrong--sign" muons produced in 
the accelerator complex upstream of the storage ring. 
Finally, Fig.~\ref{plane_tau_fig} shows the single--event contours in 
the ($\Delta m^2$, $\sin^2(2\theta)$)--plane for $\nu_e$--$\nu_\tau$ 
oscillations, where once again the search is based on looking 
for wrong-sign muons produced in charged current interactions. 
The pure $\nu_e$ component in the neutrino beam from a muon storage 
ring would enable the sensitivity of $\nu_e$--$\nu_\tau$ searches 
to approach that of $\nu_\mu$--$\nu_e$ searches, and result in an 
improvement beyond the sensitivities of past $\nu_e$--$\nu_\tau$ 
oscillation searches~\cite{tau1,tau2} by many orders of magnitude.

\section{Conclusions}

A very intense muon source of the type currently being 
developed for a future high-luminosity muon 
collider would provide sufficient muons to make very 
intense neutrino and antineutrino beams. 
If O($10^{20}$) muons per year were allowed to decay within a 
20~GeV/c storage ring with a straight section pointing in the desired 
direction, the resulting beams would produce hundreds of charged 
current neutrino interactions per year in a 10~kT detector on the 
other side of the Earth. 
High beam intensities, together with the purity of the initial 
flavor content of the neutrinos and 
antineutrinos within the beam, would 
provide unique opportunities for short-, long-, and very-long-baseline 
neutrino experiments. 
The physics program that can be pursued with 
muon storage ring neutrino sources could begin 
with a short baseline experiment using  relatively low energy 
muons (O(1~GeV)) as soon as an intense muon source became operational, 
and be extended to include higher energy 
muon storage rings and longer baseline experiments 
as higher energy muon beams became available. 

\vspace{0.5cm}

{\bf Acknowledgments} 
\vspace{0.2cm}\\
I would like to thank Carol Johnstone, Ray Stefanski, 
and Alvin Tollestrup 
for valuable comments on the manuscript. 
This work was performed at the Fermi National Accelerator 
Laboratory, which is operated by Universities Research Association,
under contract DE-AC02-76CH03000 with the U.S. Department of
Energy.

\newpage

\clearpage

\begin{table}
\renewcommand{\baselinestretch}{1}
\label{tab:com}
\centering{
\caption{Summary of the neutrino oscillation experimental configurations 
considered in 
the text. The number of $\nu_e$ charged current interactions per year 
and the mean energies of the interacting neutrinos are listed for 
a detector of mass $m_{DET}$ a distance L from a storage ring in which 
$7.5 \times 10^{20}$ 
unpolarized positive muons per year are injected with momenta p, and 
25\% of the muons decay in the straight section pointing at the experiment.}
\vspace{0.6 cm}
\begin{tabular}{cccccc}  \hline\hline
p & $m_{DET}$ & L & $<E_\nu>$ & L/$<E_\nu>$ & $\nu_e$ CC \\
(GeV/c) & (kT) & (km) & (GeV) & (km/GeV) & interactions/yr \\ \hline
20 & 10 & 9900 & 13 & 744 & $ 1 \times 10^3$ \\
20 & 10 & 732 & 13 & 57 & $2 \times 10^5$ \\
10 & 10 & 732 & 6.6 & 111 & $3 \times 10^4$ \\
1.5 & 1 & 1 & 1 & 1 & $5 \times 10^6$\\ \hline \hline
\end{tabular}
}
\end{table}

\clearpage

%
%

\end{document}